\documentclass[preprint,review,12pt]{elsarticle}

\usepackage{amssymb}
\usepackage{multirow}

\journal{Computer in Medicine and Biology}

\begin{document}

\begin{frontmatter}



\title{Cross-dataset COVID-19 Transfer Learning with Cough Detection, Cough Segmentation, and Data Augmentation}


\author{Bagus Tris Atmaja$^{a,b}$, Zanjabila$^a$, Suyanto$^a$, Akira Sasou$^b$}



\begin{abstract}
This paper addresses issues on cough-based COVID-19 detection. We propose a cross-dataset transfer learning approach to improve the performance of 
COVID-19 detection by incorporating cough detection, cough segmentation, and data augmentation. The first aimed at removing non-cough signals and cough signals with low probability. The second aimed at segregating several coughs in a waveform into individual coughs. The third aimed at increasing the number of samples for the deep learning model. These three processing blocks are important as our finding revealed a large margin of improvement relative to the baseline methods without these blocks. An ablation study is conducted to optimize hyperparameters and it was found that alpha mixup is an important factor among others in improving the model performance via this augmentation method. A summary of this study with previous studies on the same evaluation set was given to gain insights into different methods of cough-based COVID-19 detection.

\end{abstract}


\begin{highlights}
\item Three processing blocks were proposed to improve cough-based COVID-19 detection.
\item The study reports ablation studies on these three blocks and hyperparameters tuning. 
\item It also summarizes previous studies on the same test set of COVID-19 detection.
\end{highlights}


\end{frontmatter}


\section{Introduction}
\label{sec:intro}
The coronavirus disease that spread at the end of 2019 in China (COVID-19) and in early 2020 over the world is showing the unpreparedness of humans for the pandemic. Although a similar case has occurred previously (SARS, MERS), the response to diagnose the virus in a short time to prevent its spread is not optimal. The gold standard polymerase chain reaction (PCR) test takes time in days and hours. The need for preliminary screening by using other tools is crucial to avoid the spread of the virus.

The human voice is new blood. For many diseases, a blood test is the main tool to assess the general condition of the human body. By performing a blood test, the infection of an organ could be checked; the function of certain organs could be monitored. Modern medical imaging techniques (fMRI, CT-scan, X-ray) help medical doctors with accurate diagnoses. On the other side, the use of humans voice for diagnosing particular diseases is limited. Nevertheless, acoustic signals from the human body represent the state of the parts to produce these signals. The sound of the human voice is new blood that could be used for diagnosing particular diseases like in blood and imaging.

It has been evidenced that the human voice could be used as the main tool for diagnosing diseases related to voice production: pathological voice detection \cite{Gupta2016}, pertussis \cite{Pramono2016}, asthma \cite{khassaweneh2013}, and respiratory diseases \cite{Swarnkar2013}. Moreover, the use of acoustic analysis has been proven effective for Alzheimer's disease classification (accuracy of 93.30\%) \cite{Bertini2021} and lung disease (accuracy of 98.92\%) \cite{Shuvo2021}. These examples show the potency of the human voice for diagnosing diseases, particularly voice-related diseases.

Attempts to explore acoustic analysis beyond voice-related diseases have been conducted for COVID-19 detection. Several modalities have been tried including cough, breathing sounds, and speech or voice (including vowels). The first two modalities contain indicators for the symptoms of COVID-19: continuous cough and shortness of breath. We choose the first modality due to the large available cough data. The cough sounds also showed the highest specificity among other modalities in the previous study \cite{Han2022}.

\begin{figure}[htbp]
    \centering
    \includegraphics[width=0.85\textwidth]{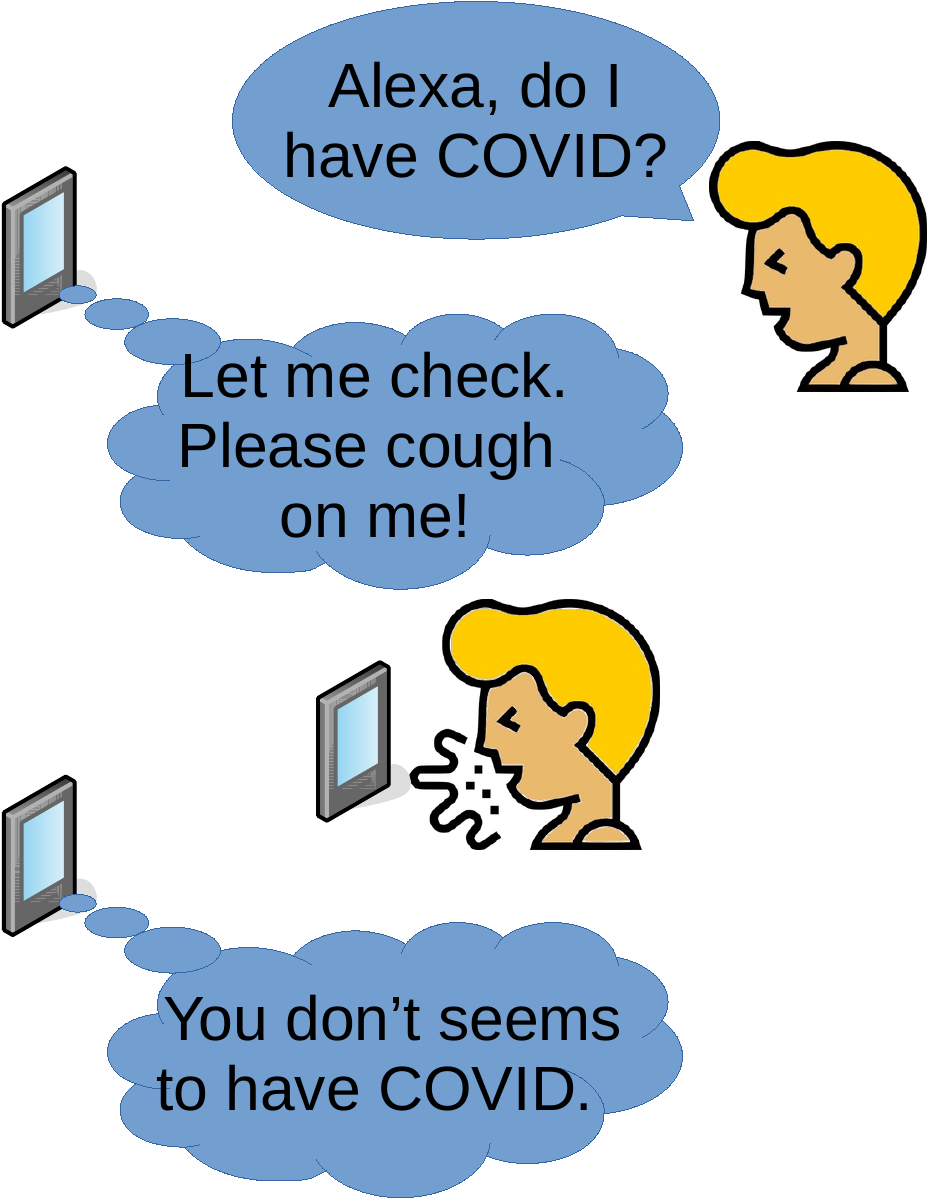}
    \caption{Illustration of the smartphone application for COVID-19 detection.}
    \label{fig:app}
\end{figure}

Nowadays, computers are part of human daily equipment, ranging from personal computers to smartphones, which could be used for such acoustic analysis. Given the benefit of diagnosing particular diseases (COVID-19, dementia, depression), the development of an application on a smartphone to detect flare-ups of pulmonary disease \cite{Anthes2020}. In the future, we predict that smartphones could be furnished with acoustic analysis applications to detect diseases like COVID-19 (Figure \ref{fig:app}). On that day, we could simply ask our smartphone if we have COVID-19 or not for preliminary screening, as in Figure \ref{fig:app}. This application will likely be available not only for COVID-19 but also for other diseases and disorders \cite{Anthes2020}. In the current stage, applications for classifying the crying of a baby, counting coughs, and detecting anxiety and depression are available in the market. All of these applications are based on acoustic analysis. 

To speed up research on cough-based COVID-19 detection, we contribute three 
important processing blocks for improving the generalization capability of 
COVID-19 detection via transfer learning and cross-dataset evaluation. These 
blocks are cough detection, cough segmentation, and data augmentation. In 
addition to the main contribution above, we also performed an ablation study to 
optimize the hyperparameters in the transfer learning stage and a summary of 
this study with previous studies on the same evaluation (test) dataset.


\section{Related work}
Research on voice-based COVID-19 detection have been conducted since the spread of the coronavirus disease. Many researchers have proposed different approaches to classify human voices into healthy and unhealthy classes for COVID-19 detection, including the use of cough sounds. These approaches could be classified into acoustic features related to COVID-19, ensemble methods, transfer learning, and multimodal analysis.

The search of voice biomarkers related to COVID-19 have been attempted in \cite{Quatieri2020,Islam2022,Vahedian-azimi2021,Bartl-Pokorny2021a}. In \cite{Quatieri2020}, the authors proposed a framework for COVID-19 biomarkers based on the coordination of speech-production subsystems. Motivated by a unique feature of COVID-19 involving lower and upper respiratory tract inflammation, the authors measure Cohen's effect size between pre- and post, laryngeal motion (via fundamental frequency and cepstral peak prominence), and coordination of laryngeal and articulatory (via the center of formants). The results show reduce of complexity (measured in effect size) between pre- and post-COVID-19 on interview-like voices. This result indicates a possible biomarker for COVID-19 detection via respiratory function captured by acoustic signals. Other studies \cite{Islam2022,Vahedian-azimi2021,Bartl-Pokorny2021a} evaluated the existing acoustic features for specific model. For instance, the authors of \cite{Vahedian-azimi2021} measured the F-score of 25 acoustic features and found that maximum phonation time (MPT) is the most important feature in their study with /a/ vowel sounds.

The use of ensemble methods for COVID-19 detection has been proposed in \cite{Mohammed2021,Chowdhury2022,Casanova2021}. In \cite{Mohammed2021}, the authors proposed an ensemble of CNN classifiers from different acoustic features. In \cite{Chowdhury2022}, the authors proposed an ensemble-based multi-criteria decision making (MCDM). These criteria include accuracy, AUC, precision, recall, F1-score, sensitivity, and specificity; the MCDM considers several criteria instead of one. In \cite{Casanova2021}, the authors evaluated ensemble methods using different seed numbers. All the reported results show that ensemble methods can improve the performance of COVID-19 detection.

Transfer learning now is gaining popularity in many fields, and has been experimented with audio classification for COVID-19 detection. In \cite{Casanova2021}, the authors transfer the knowledge of the pre-trained model from the AudioSet dataset to ComParE-CCS dataset. In \cite{Han2022}, the authors transfer the knowledge learned from a large audio dataset to the Cambridge COVID-19 sound dataset. Both studies revealed the effectiveness of transfer learning to improve the representation of cough sounds in COVID-19 detection.

The use of multimodal (i.e., cough, vowel, and breathing sounds) in COVID-19 detection has been evaluated in \cite{Sharma2022a,Han2022}. In \cite{Sharma2020}, the authors extracted local binary patterns and Haralick's features from the spectrogram to analyze the audio textural behavior of cough, breath, and speech sounds. The authors achieved accuracy rates of 98.9\% for 5-class and 72.2\% for 2-class classification. The authors of \cite{Han2022} reported that a multimodal approach outperformed any single modality approach; breathing achieved the best sensitivity and cough sounds achieved the best specificity in unimodal evaluations.

Instead of choosing those approaches, we choose to evaluate different pre-processing blocks for feature extraction. In automatic speech recognition and speech emotion recognition, a little step in pre-processing is important, e.g., extracting silence region \cite{Wang2008,Atmaja2020f}. In this study, we proposed two techniques related to cough and evaluated a technique related to general machine learning. Two techniques related to cough are cough detection and cough segmentation. Detecting cough will filter out non-cough signals and cough signals with low probabilities. Segmenting cough will split several coughs in a waveform into individual coughs. We argued that this splitting method is more useful than fixed-time splitting (e.g., in \cite{Mohammed2021}) since the region of cough is extracted based on acoustics. The last block is data augmentation to increase the number of samples after cough detection and cough segmentation.

\section{Methods}
\begin{figure}[htbp]
    \centering
    \includegraphics[width=0.8\textwidth]{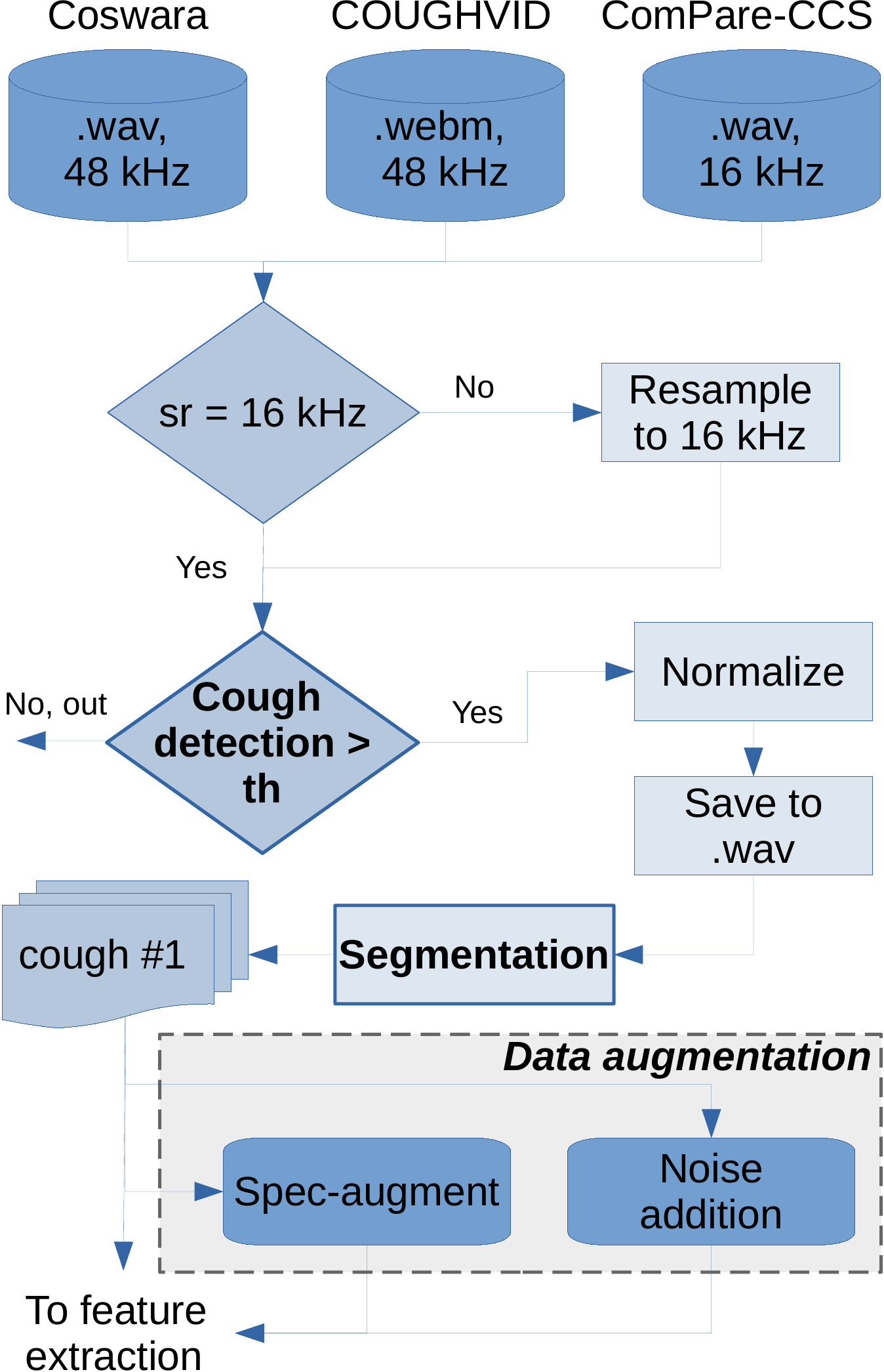}
    \caption{The flowchart for the proposed method; three processing blocks are proposed to improve cough-based COVID-19 detection: cough detection, segmentation, and data augmentation.}
    \label{fig:flowchart}
\end{figure}

This method used in this study is based on the previous work \cite{Casanova2021}. That work only obtained a slight improvement from the original baseline dataset \cite{Schuller2021}, although the authors have employed advanced techniques by utilizing transfer learning, data augmentation, and ensemble methods. We proposed to evaluted three processing blocks to improve the previous results: cough detection, cough segmentation, and data augmentation. The method is shown in Figure \ref{fig:flowchart}. Nevertheless, we describe the dataset, classifier, and evaluation metric for a complete understanding of the method.

\subsection{Datasets}
Three datasets are merged to enable cross-dataset evaluations. The test set is taken from ComParE-CCS 2021 test set for comparing our method with previous studies. The rest of ComParE-CCS dataset is merged with Coswara and COUGHVID datasets. The description of each dataset is shown below.

\noindent Coswara \\
Coswara is a database of breathing, cough, and voice sounds for COVID-19 diagnosis \cite{Sharma2020}. Among the three types of sound (speech, cough, and breathing), we only used the cough sounds. These sounds are recorded through worldwide crowdsourcing using a website application. Although the recording is targeted at worldwide participants, about 80\% of the speakers came from India. We resampled the original audio files from 48 kHz to 16 kHz. The cough sounds are then manually normalized in a range [-1, 1] and saved again in a WAV format (Figure \ref{fig:flowchart}). The normalization process was conducted by using the librosa \cite{McFee2020} toolkit. The Coswara dataset is retrieved from their GitHub repository with commit ID ``401b516". 

\noindent COUGHVID \\
COUGHVID is a crowdsourcing dataset for the study of large-scale cough analysis algorithms \cite{Orlandic2021}. The dataset focuses on three features: detection of cough, expert labeling of cough, and high correlation between symptomatic and COVID-19 labels and location of the speakers with high infection rates. We employed both the cough detection model and COUGHVID data in this cough-based COVID-19 detection study. The original sampling rate of 48 kHz was resampled to 16 kHz. The format is also converted from WEBM to WAV  with normalization in a range of [-1, 1]. The COUGHVID dataset was retrieved from their Zenodo repository with Version 3.0.

\noindent ComParE-CCS\\
ComParE-CCS is a dataset for cough-based COVID-19 detection \cite{Schuller2021} originally for the INTERSPEECH 2021 Computational Paralinguistic Challenge (ComParE). One of the sub-challenges is the COVID-19 cough sub-challenge (CCS) and COVID-19 speech sub-challenge (CSS). We only used data from CCS to merge with the previous aforementioned datasets. The original dataset contains 725 recordings divided into Train (286 samples), Development (231 samples), and Test (208 samples). To enable comparison with previous studies, we used the same test partition. The other partitions (Train and Development) are merged with the previous datasets (Coswara and COUGHVID). The sampling rate was kept at 16 kHz, but the audio files are normalized in a range [-1, 1] following the same treatment on the previous datasets. This ComParE-CCS dataset was obtained by emailing the organizers of the challenge.

The summary of the number of samples for all datasets is shown in Table \ref{tab:dataset}. Note that these numbers are after cough detection and segmentation processing blocks. There is a decrease in the number due to filtering by the cough detection method. For instance, the original ComParE-CCS train, development, and test sets are 286, 231, and 208 samples, respectively. After the cough detection and segmentation, the numbers of samples are 236, 134, and 154 samples, respectively. The same thing happens to the Coswara and COUGHVID datasets. 

\begin{table}[htbp]
    \caption{Number of samples for the final dataset (after cough detection and cough segmentation)}
    \centering\begin{tabular}{l c c c c}
    \hline 
    Dataset & Train & Devel. & Test & Total \\
    \hline
    Coswara & 564 & 100 & - & 664 \\
    COUGHVID & 712 & 126 & - & 838 \\
    ComParE-CCS & 236 & 134 & 154 & 524 \\
    \hline
    Total & 1512 & 360 & 154 & 2026 \\
    \hline    
    \end{tabular}
    \label{tab:dataset}
\end{table}

\subsection{Acoustics Features}
Mel spectrogram is a powerful acoustic feature that is widely used in
automatic speech recognition \cite{Guo2019}, language identification
\cite{Choi2021}, speech emotion recognition \cite{Liu2019}, music tagging
\cite{Choi2018}, and cough-based COVID-19 detection \cite{Mohammed2021}.
Borrowing the success of mel spectrogram, we employed the log mel spectrogram
as the acoustic feature for this study. The log mel spectrogram is computed by
using the TorchAudio \cite{Yang2022} toolkit. The parameters are set as
follows: n\_fft=1024, hop\_length=320, n\_mels=64, mel\_fmin=0, and
win\_length=1024. 

\begin{figure}[htbp]
    \centering
    \includegraphics[width=0.7\textwidth]{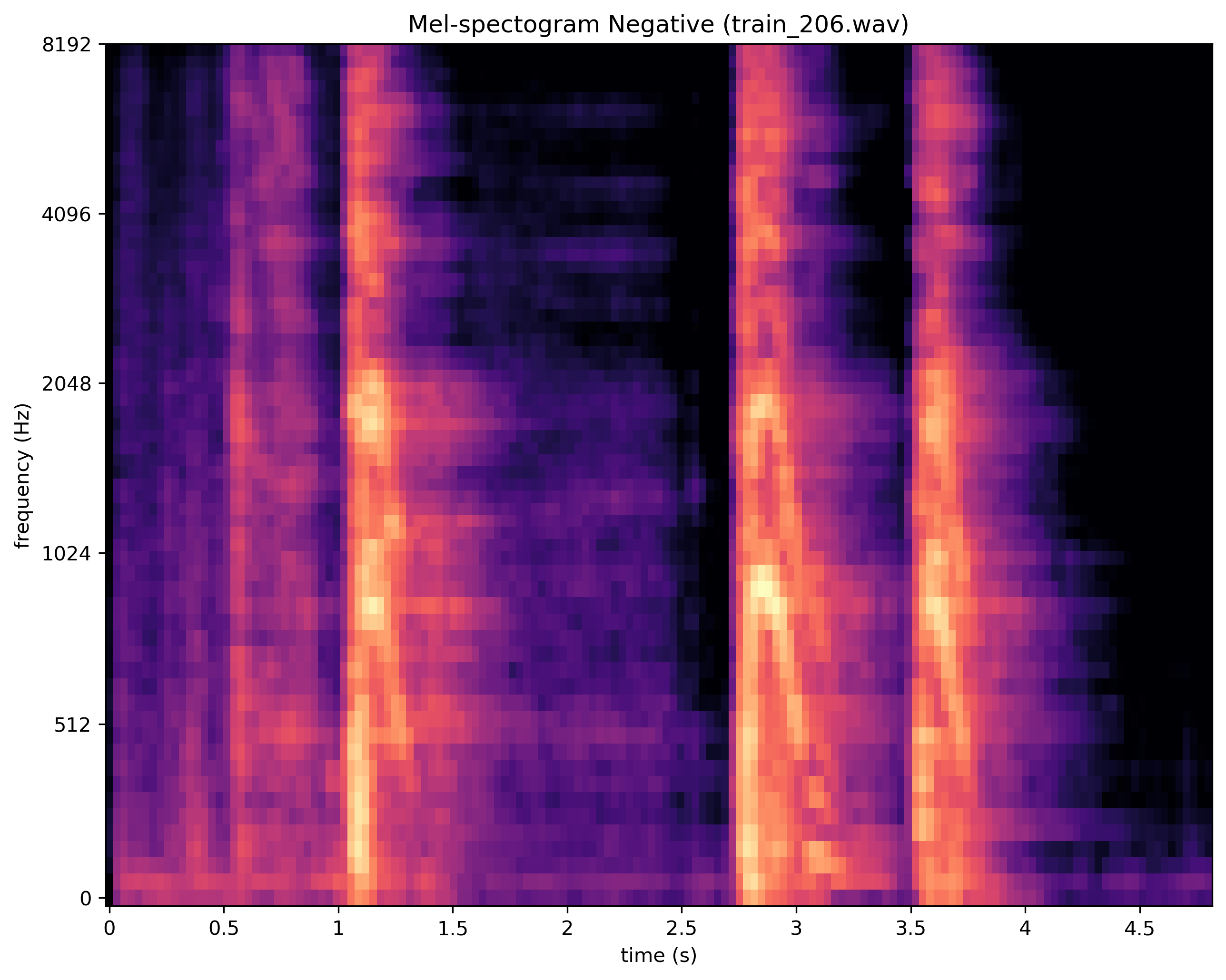}
    \caption{Mel spectrogram of the cough sound}
    \label{fig:melspec}
\end{figure}

A mel spectrogram is a spectrogram in the mel scale. Suppose an audio spectrogram (Short Time Fourier Transform, STFT) with a center frequency $f$ (Hertz, Hz; half of a sampling rate). The mel-scale conversion from Hertz is then given by, 

\begin{equation}
    m = 2595 \log_{10} \left( 1 + \frac{f}{700} \right).
\end{equation}
The visualization of mel spectrogram formulated above is shown in Figure \ref{fig:melspec}, which is taken from the train data in the ComParE-CCS dataset.

Since the pre-trained model, PANNS \cite{Kong2020}, is trained in log mel spectrogram, we convert the mel spectrogram to log mel spectrogram by using the following equation:
\begin{equation}
    log~S = 20 \times log_{10} \left( \frac{S}{ref} \right), 
\end{equation}
where S is the spectrogram in a mel scale (mel spectrogram) and $ref$ is the reference value. The reference value is set to 1.0. 

\subsection{Transfer learning and classifiers}
\begin{figure}[htbp]
    \centering
    \includegraphics[width=0.85\textwidth]{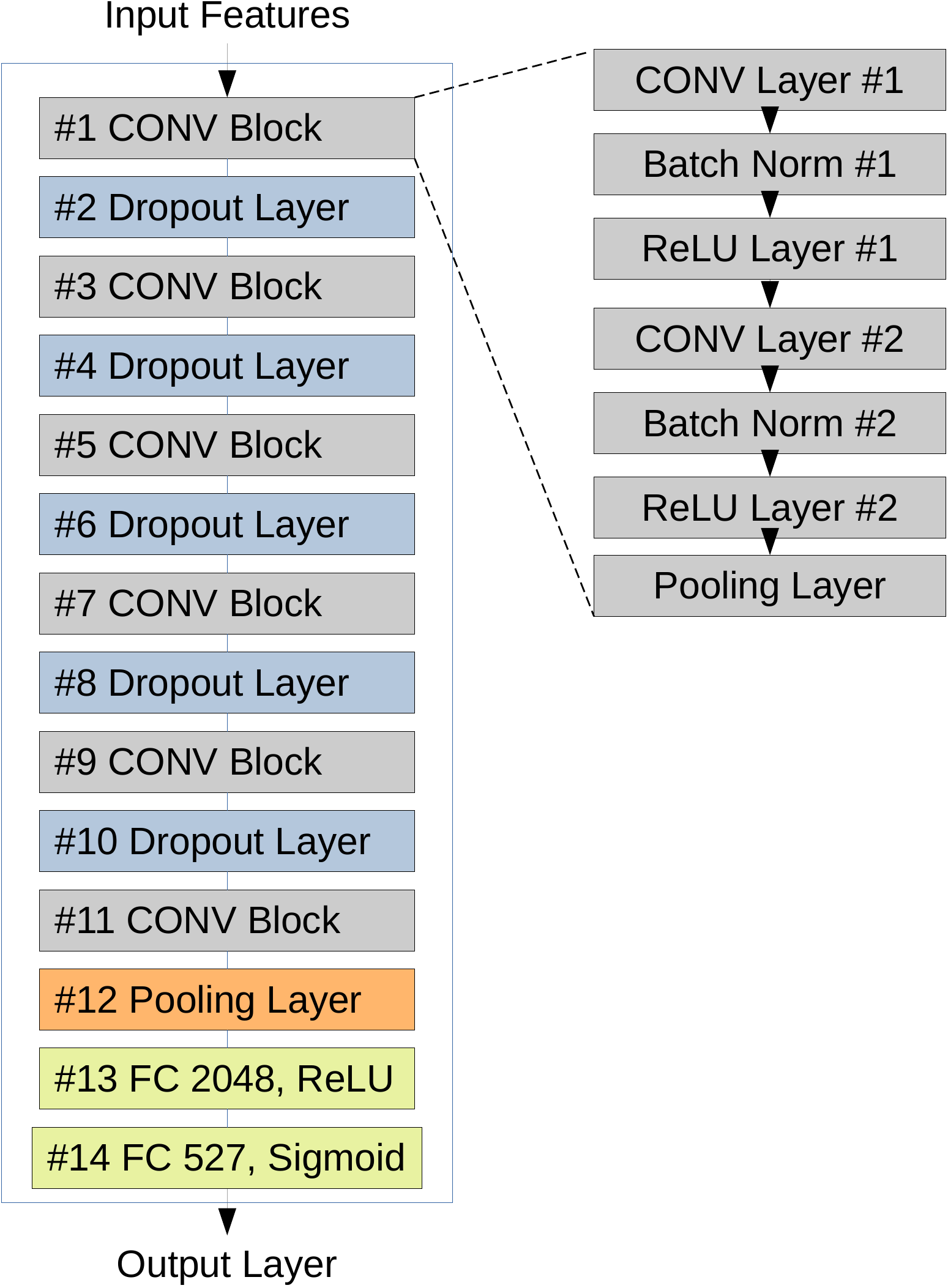}
    \caption{Architecture of the CNN14 for transfer learning; the last embedding (2048-dims) is used as the input of transfer learning with a linear layer with 2048-node input and 2-node output.}
    \label{fig:cnn14}
\end{figure}

We employed a transfer learning from the previous pre-trained model for the AudioSet dataset (527 classes) \cite{Kong2020}. The PANNS (pre-trained audio neural networks) outperformed the previous models on the original AudioSet task and other tasks (transfer learning), including ESC-50 (50 sound events), DCASE 2019 Task 1 (Acoustic scene of 12 cities), DCASE 2028 Task 2 (automatic audio tagging of 41 labels), and GTZAN (music genre classification with 10 genres). We used PANNs with CNN14 architecture shown in Figure \ref{fig:cnn14}. The model is trained with the AdamW optimizer \cite{Loshchilov2017} with a learning rate of 0.001, weight decay of 0.01, and batch size of 16. The model is trained for 100 epochs without early stopping. 

The input of the classifier is log mel spectrogram and the output is binary classification (two classes) for positive and negative prediction. This output is a transfer learning from AudioSet with 527 classes to two COVID-19 classes via a linear layer (nodes = 2048). The pre-trained model is used as a feature extractor to obtain the embedding in the last layer (2048-dims). The output layer (in transfer learning) is activated by using a sigmoid function. The output is then either positive (patient, ``1") or negative (control, ``0"). 

\subsection{Cough detection}
The goal of the cough detection processing block is to filter out non-cough signals and cough signals with low probability (low cough recognition rate). We employed a model from the previous study \cite{Orlandic2021} by utilizing XGBoost classifier \cite{Chen}. The performance of the model with threshold=0.8 is 88.1\% of balanced accuracy (unweighted accuracy). There are 18 types of acoustic features for the cough detection model, from MFCC to spectral decrease. We analyze the distribution of each dataset (recognition rate of cough) and evaluate different thresholds of cough detection for cough-based COVID-19 classification in the experiments. 

\subsection{Cough segmentation}
The goal of cough segmentation is to segment a single audio file containing several coughs into several audio files (or arrays) containing one cough each. 
The implementation of this cough segmentation is based on a previous study \cite{Atmajab}. There are two segmentation methods evaluated in this study, hysteresis comparator and RMS threshold. In the former, we define two boundaries (upper and lower) and detect the signal with rapid spikes in power \cite{Orlandic2021}. In the latter method, we use a single threshold; if the signal is above the threshold, it is considered a cough. The threshold is set to 0.09 after normalization. 

\subsection{Data augmentation}
Data augmentation is a technique to increase the size of the dataset by generating new samples from the existing samples. The goal of data augmentation is to increase the size of the dataset and reduce the overfitting problem. We employed the following data augmentation techniques: mixup \cite{Zhang2018}, SpecAugment \cite{Park2019}, and noise addition. The noise addition is performed by convolving the original audio data with MUSAN noise corpus \cite{Snyder2015a}. The MUSAN dataset contains three types of noises: speech, music, and noise; we only chose noise type to add to the original audio data. The noise is added to the original audio data with a random SNR (signal-to-noise ratio) between 0 and 15 dB.

While mixup augmentation is used in the baseline, we evaluate the effectiveness of adding data with SpecAugment, noise addition, and a combination of both. A further evaluation for mixup augmentation is performed by evaluating different alpha mixup values ranging from 0.1 to 1.0 with a 0.1 step. This last evaluation is also performed to tune other hyperparameters: learning rate and weight decay.

\subsection{Evaluation metric}
We use a single metric namely unweighted accuracy (UA) as the evaluation metric. The UA is defined as the number of correctly classified samples divided by the total number of samples. The UA is a common metric used in previous studies \cite{Schuller2021,Casanova2021}. Using this metric enables us to compare the performance of the proposed model with the previous studies on the same test set (ComParE-CCS). This metric is also known as balanced accuracy or unweighted average recall (UAR).

\section{Results and discussion}
We present our results and discuss them in six parts: data exploration, cough detection, cough segmentation, data augmentation, ablation study, and a summary of previous studies. These parts are described in the following subsections.

\subsection{Data exploration}
\begin{figure}[htbp]
    \centering
    \includegraphics[width=0.85\textwidth]{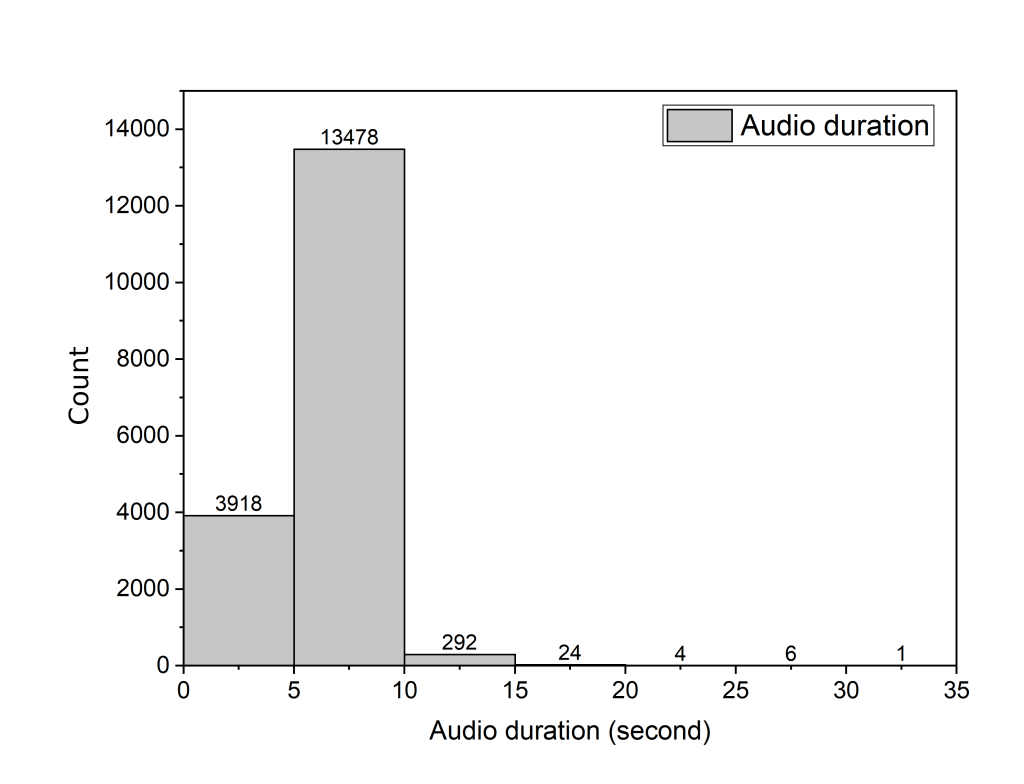}    
    \caption{Distribution of data based on duration}
    \label{fig:audio_duration}
\end{figure}

\begin{figure}[htbp]
    \centering
    \includegraphics[width=0.85\textwidth]{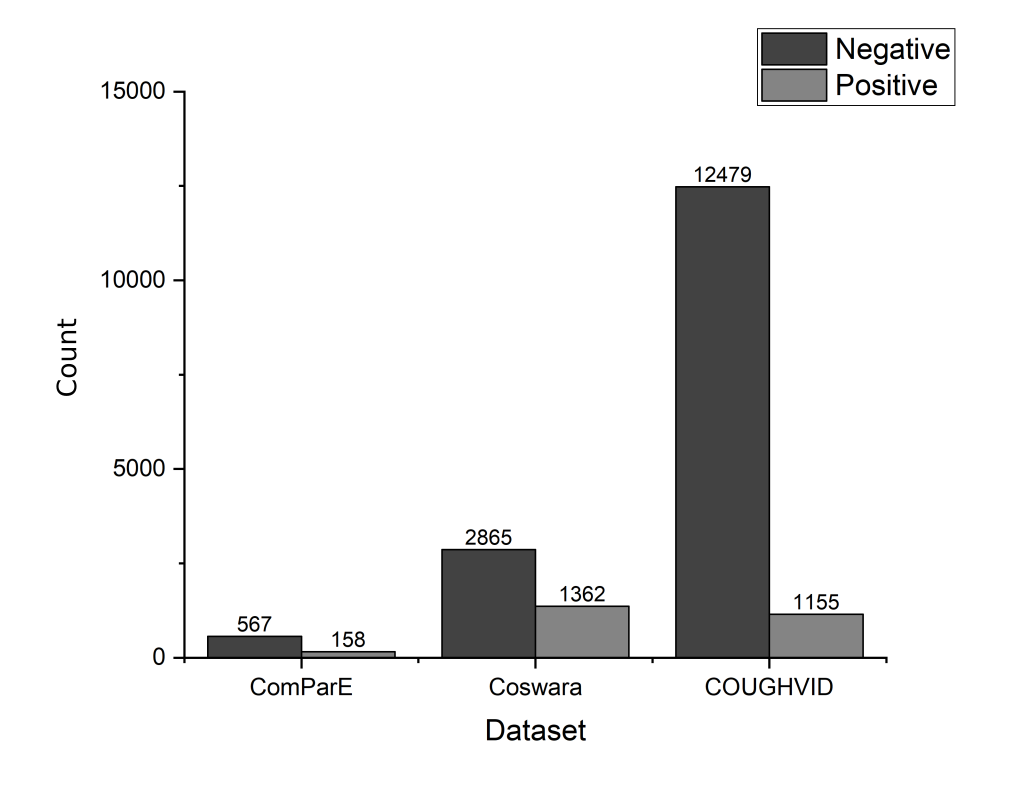}    
    \caption{Data distribution of positive and negative samples on the original three datasets}
    \label{fig:data_distribution}
\end{figure}

At first, it is necessary to explore the data to gain insights about information about them. We plot the duration of cough sounds in Figure \ref{fig:audio_duration} to know the length of typical cough files. The majority of cough sounds have a duration of between 5-10 seconds. Another data exploration is the distribution between positive and negative samples. Figure \ref{fig:data_distribution} shows the distribution of positive and negative samples in the original three datasets. Naturally, the negative samples (healthy cough) are more than the positive samples (COVID-19 cough), as shown in the figure. The use of negative cough may affect the classification of cough-based COVID-19 detection since some cough sounds in negative samples are not natural (i.e., artificial coughs).

The next data exploration is to find the best ratio for splitting training and development data. While the test data is fixed, the portion of data for training and development can be varied, e.g., 80:20, 70:30, and 60:40. Reference \cite{Gholamy2018} argued that the best division for training and test is 80:20, while V.R. Joseph \cite{Joseph2022} proposed portion of $\sqrt{p}:1$ for linear regression where $p$ is the number of parameters in linear regression. Another strategy for splitting data is as proposed by I. Guyon \cite{Guyon1997} based on the complexities of data. For instance, the larger data, the smaller amount of validation set should be reserved than the smaller data. 

\begin{figure}[htbp]
    \centering
    \includegraphics[width=0.85\textwidth]{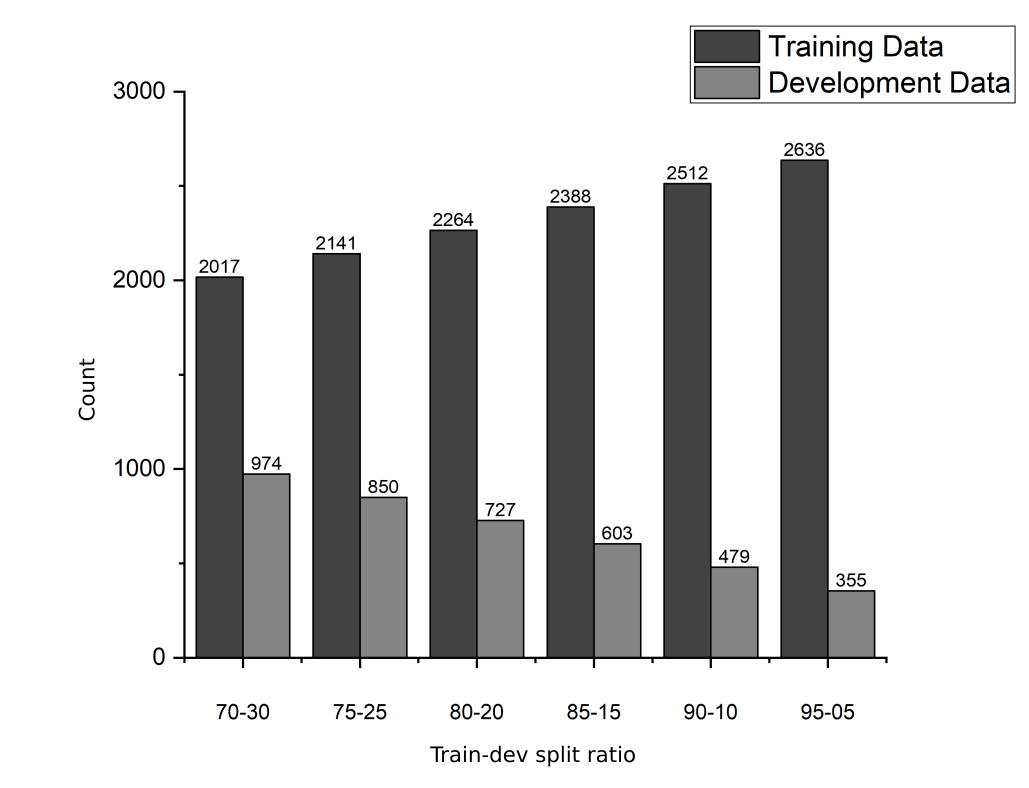}    
    \caption{Count of data on different train-test split ratio}
    \label{fig:train_test_split}
\end{figure}

In this study, we evaluate six different ratios for splitting training and validation data. The different number of data for each split ratio is given in the Figure \ref{fig:train_test_split}. The results are shown in Table \ref{tab:split_ratio}. The optimal value is given by a ratio of 85:15 for training:validation. Note that this ratio contains positive samples only for Coswara and COUGHVID datasets, according to our best experiments. For ComParE-CCS, the ratio includes negative samples that will be used for balancing other datasets. The next experiments are conducted using this 85:15 split ratio.

\begin{table}[htbp]
    \caption{UA of different split-test ratio}
    \centering\begin{tabular}{c c c c}
        \hline \multirow{2}{*}{ Variation } & \multicolumn{2}{c}{ Ratio } & \multirow{2}{*}{ UA }\\
         & Train & Devel &  \\
        \hline
        1 & $70 \%$ & $30 \%$ & $65.18 \%$ \\
        2 & $75 \%$ & $25 \%$ & $67.65 \%$ \\
        3 & $80 \%$ & $20 \%$ & $68.14 \%$ \\
        4 & $85 \%$ & $15 \%$ & \textbf{72.68}\% \\
        5 & $90 \%$ & $10 \%$ & $68.54 \%$ \\
        6 & $95 \%$ & $5 \%$ & $69.14 \%$ \\
        \hline
        \end{tabular}
    \label{tab:split_ratio}
\end{table}

\subsection{Effect of cough detection}
\begin{figure}[htbp]
    \centering
    \includegraphics[width=0.85\textwidth]{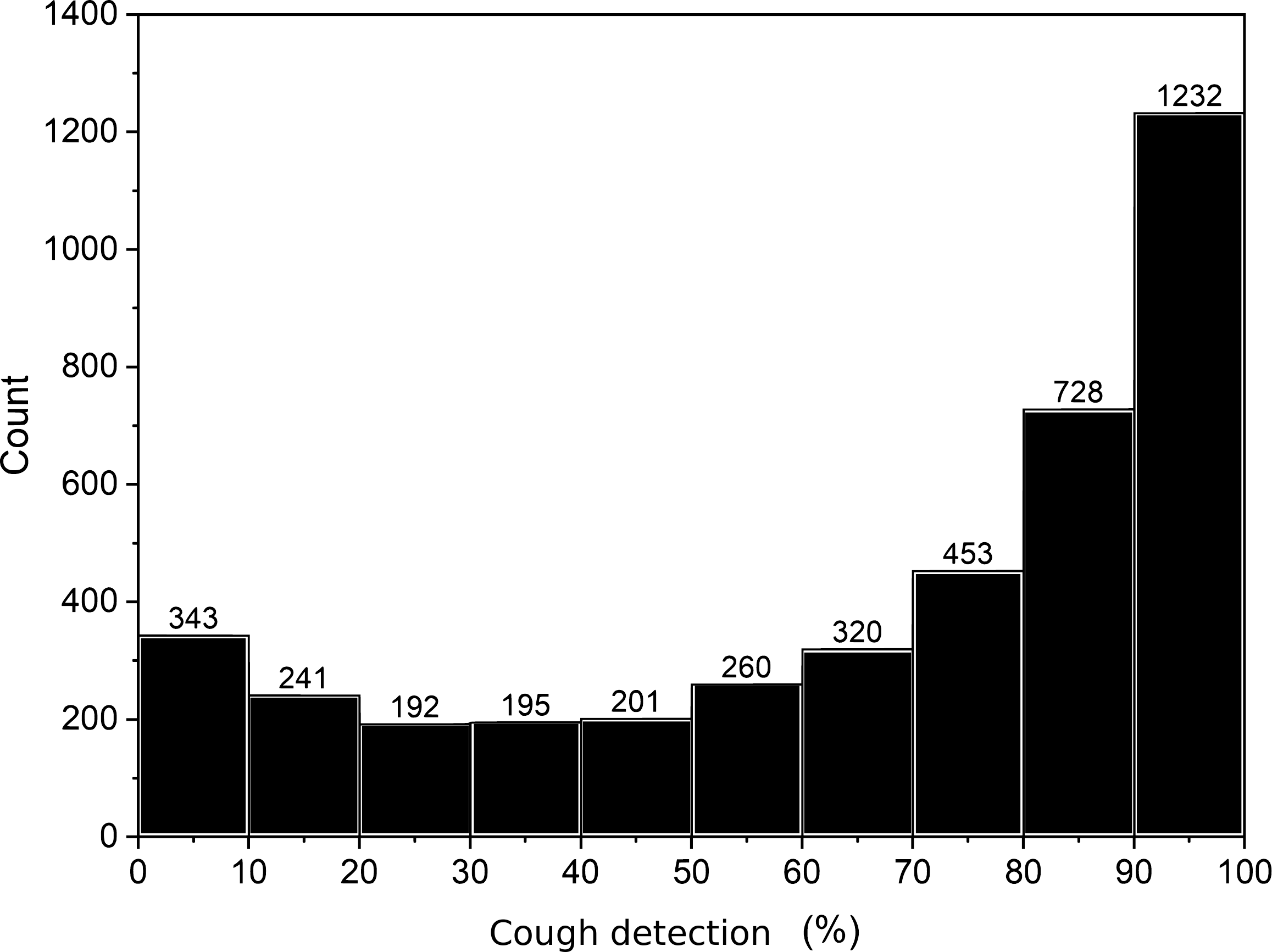}    
    \caption{Count of data for Coswara dataset on different cough detection thresholds}
    \label{fig:coswara_hist}
\end{figure}

\begin{figure}[htbp]
    \centering
    \includegraphics[width=0.85\textwidth]{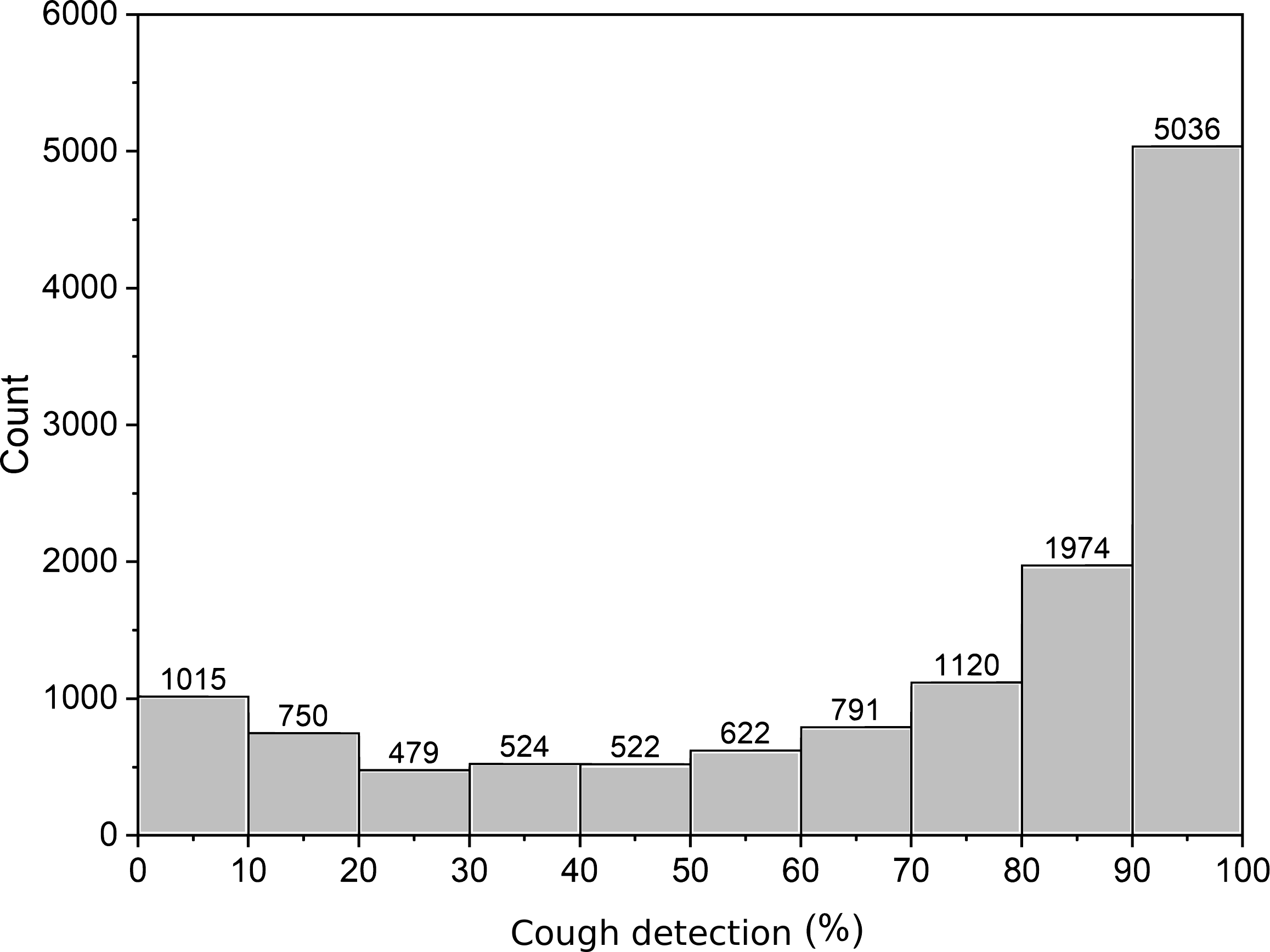}    
    \caption{Count of data for COUGHVID dataset on different cough detection thresholds}
    \label{fig:coughvid_hist}
\end{figure}

\begin{figure}[htbp]
    \centering
    \includegraphics[width=0.85\textwidth]{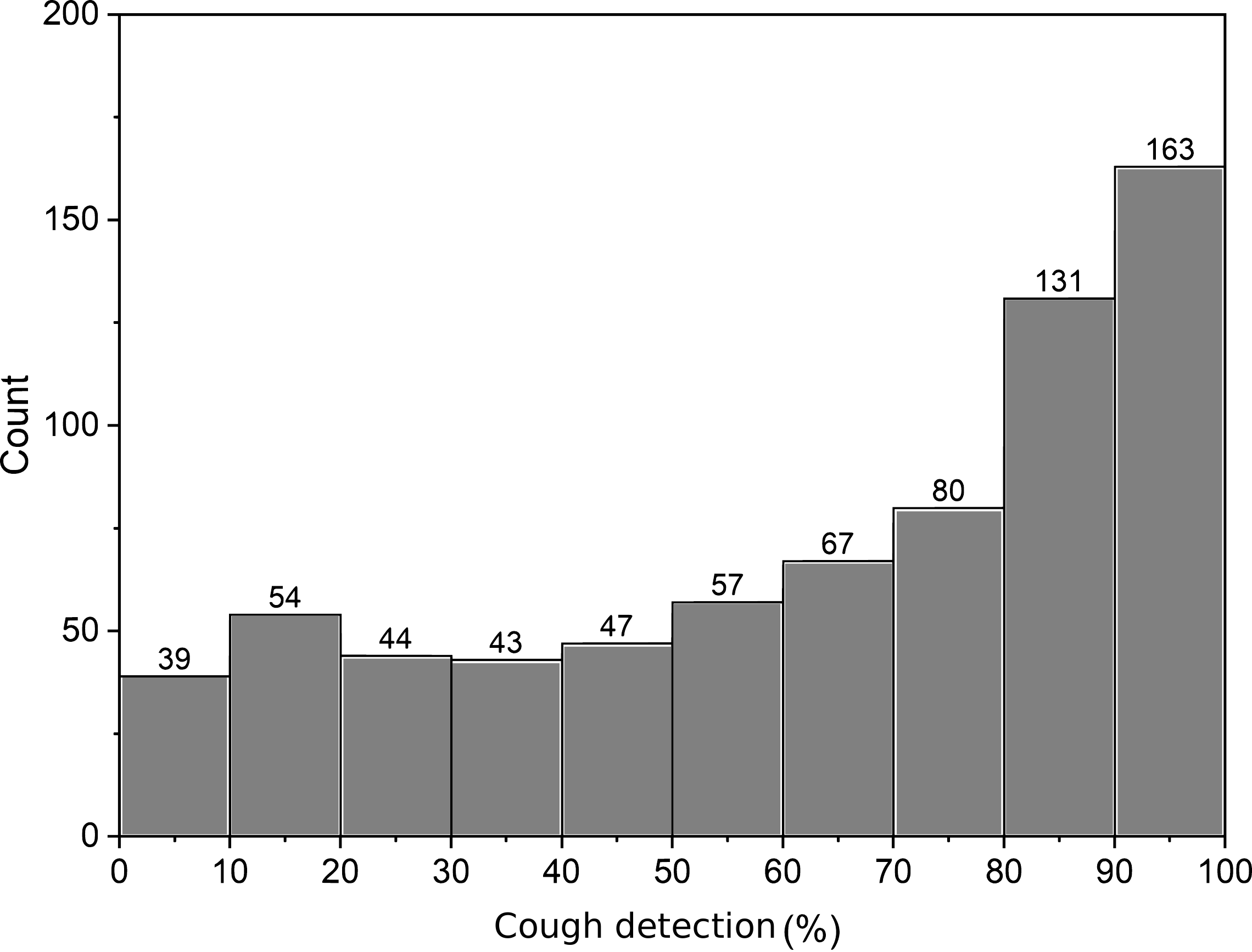}    
    \caption{Count of data for ComParE-CCS dataset on different cough detection thresholds}
    \label{fig:compare_hist}
\end{figure}

Our first proposal for enhancing the performance of cough-based COVID-19 detection is by detecting cough sounds through specific thresholds. In this block, we want to filter cough sounds with a high recognition rate only (removing low-rate cough sounds, including non-cough sounds). The model used to detect cough is based on the XGB classifier \cite{Orlandic2021}. We present distribution of recognition rate for each dataset in Figures \ref{fig:coswara_hist},\ref{fig:coughvid_hist}, and \ref{fig:compare_hist}. It has been shown that there is a high number of data with a low recognition rate. For instance, in the COUGHVID dataset, there are 1015 data with a recognition rate below 10\%. Removing low recognition rate cough sounds may improve the performance of cough-based COVID-19 detection.

\begin{figure}[htbp]
    \centering
    \includegraphics[width=0.85\textwidth]{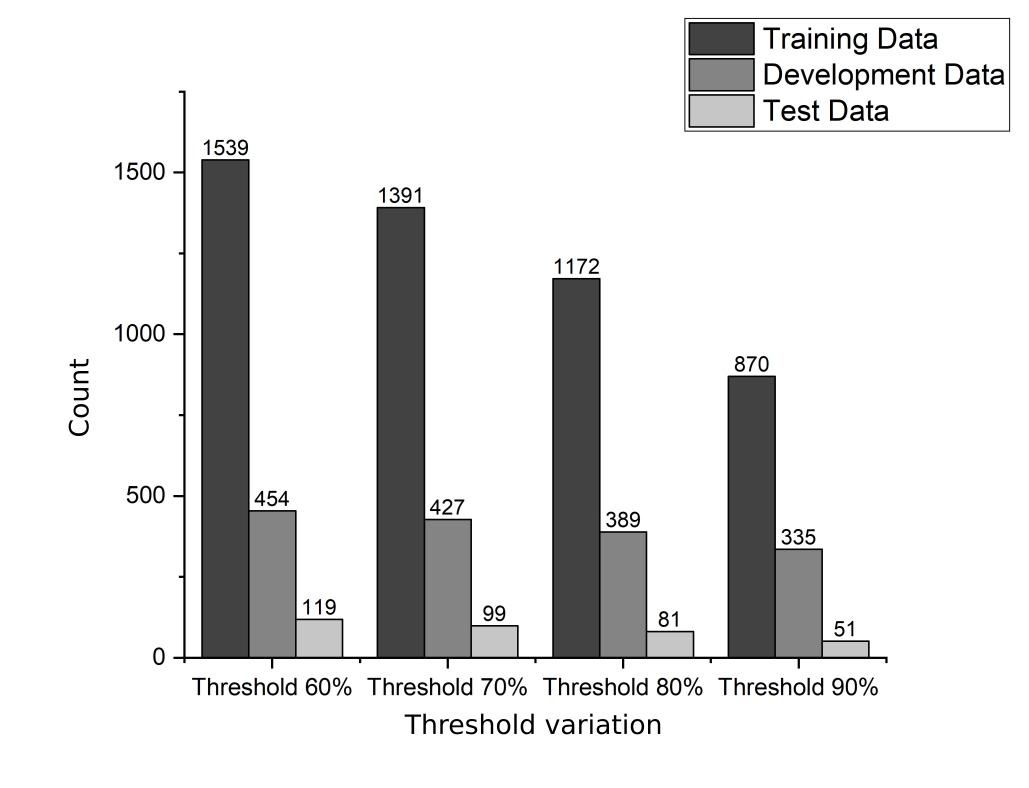}
    \caption{Count of all data on different cough detection thresholds}
    \label{fig:threshold_var}
\end{figure}

We varied the thresholds to filter cough sounds from three datasets on values of 60\%, 70\%, 80\%, and 90\%. The distribution (data count) for each threshold is shown in Figure \ref{fig:threshold_var}. Table \ref{tab:threshold_var} shows the UA of each threshold. The results show that the best threshold is 90\% for all three datasets with a UA of 72.54\%. It also shows that the smaller threshold leads to a lower UA, highlighting the importance of (our proposal by) filtering cough sounds with cough detection for COVID-19 detection. The next experiments are conducted using this 90\% threshold cough detection.

\begin{table}[htbp]
    \centering
    \caption{UA results on different cough detection thresholds}
    \begin{tabular}{ccc}
        \hline Threshold & Data count & UA \\
        \hline $60 \%$ & $2201$ & $71.40 \%$ \\
        $70 \%$ & $2026$ & $71.99 \%$ \\
        $80 \%$ & $1769$ & $72.48 \%$ \\
        $90 \%$ & $1413$ & \textbf{75.54\%} \\
        \hline
        \end{tabular}
    \label{tab:threshold_var}
\end{table}

\subsection{Effect of cough segmentation}
\begin{figure}[htbp]
    \centering
    \includegraphics[width=0.85\textwidth]{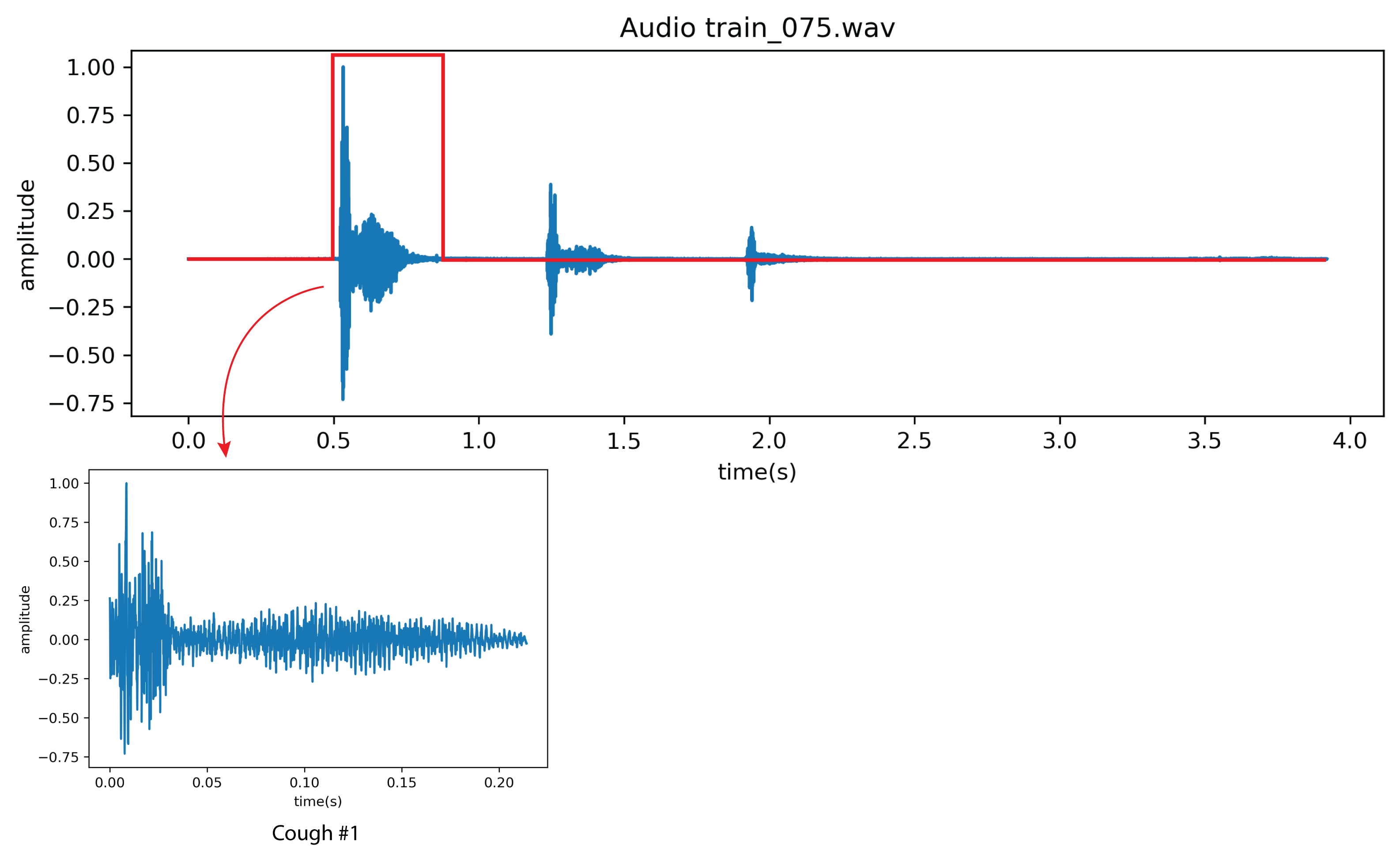}
    \caption{Segmentation of coughs using hysteresis comparator}
    \label{fig:seg_hc}
\end{figure}

\begin{figure}[htbp]
    \centering
    \includegraphics[width=0.85\textwidth]{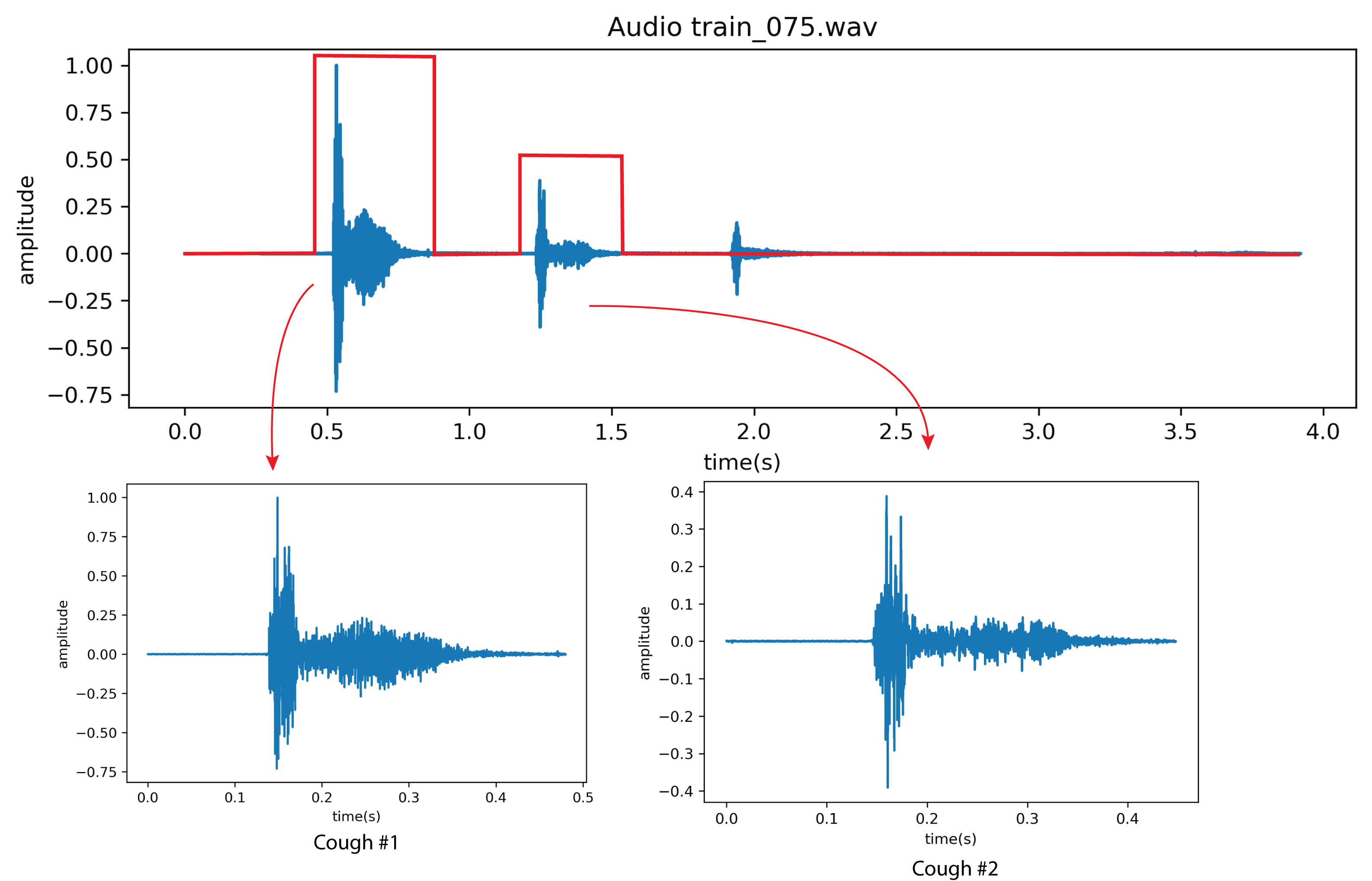}   
    \caption{Segmentation of coughs using RMS method}
    \label{fig:seg_rms}
\end{figure}

Segmenting an audio file containing several coughs into several audio files containing one cough is another proposal for enhancing COVID-19 detection. In this case, we evaluated two different algorithms to segment cough sounds. The first algorithm is based on a hysteresis comparator which compares whether a signal is within the region of cough (two thresholds) \cite{Orlandic2021}. The second algorithm is with a single threshold using RMS value \cite{Atmajab,Atmaja2020f}. Figures \ref{fig:seg_hc} and \ref{fig:seg_rms} show the segmentation results by the two algorithms from the same input file. Notice the different number of segmented files for each algorithm. The first algorithm produces a single cough, while the second algorithm produces two coughs. 

\begin{figure}[htbp]
    \centering
    \includegraphics[width=0.85\textwidth]{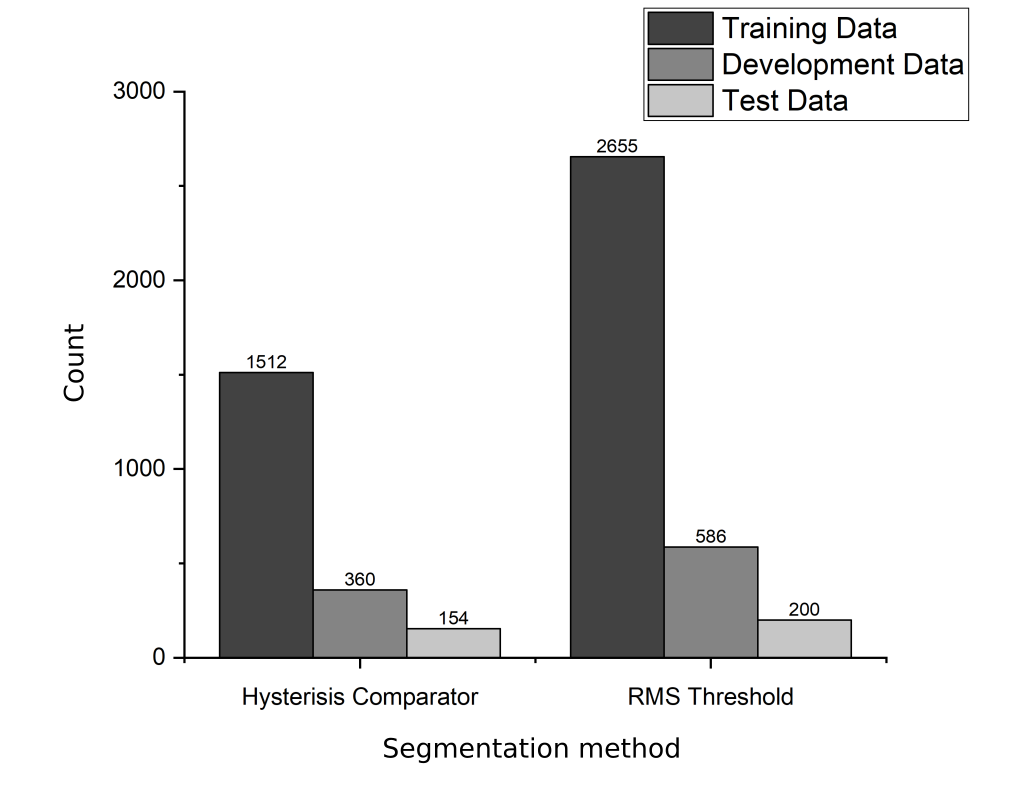}
    \caption{Count of data on different cough segmentation thresholds}
    \label{fig:seg_var_count}
\end{figure}

Figure \ref{fig:seg_var_count} and Table \ref{tab:seg_var} show the distribution (number) of new audio files after segmentation methods. The RMS method produces more segmented files than the hysteresis comparator. The UA of the two methods is shown in Table \ref{tab:seg_var}. The UA of the hysteresis comparator is 81.19\%, while the RMS method is 79.06\%. The UA of the hysteresis comparator is higher than the RMS method. The hysteresis comparator is chosen as the segmentation method for the next experiments. Compared to the previous result (Table \ref{tab:threshold_var}), both segmentation methods improve the UA without segmentation. The improvements can be seen as a factor of segmentation that standardizes the cough sound (into a single cough) and enlarges the number of samples (e.g., training data from 870 to 1512 with a hysteresis comparator). The RMS method with larger data obtained a lower hysteresis comparator method; the RMS method may be less sensitive than the hysteresis comparator on segmenting cough sounds. The next reported experiments are conducted using the hysteresis comparator segmentation method.

\begin{table}[htbp]
    \centering
    \caption{UA results on different cough segmentation thresholds}
    \begin{tabular}{c c c }
        \hline Method & Number of data & UA \\
        \hline
        Hysteresis Comparator & $2026$ & \textbf{83.19\%} \\
        RMS Threshold & $3441$ & $79.06 \%$ \\
        \hline
        \end{tabular}
    \label{tab:seg_var}
\end{table}

\subsection{Effect of data augmentation}
The last proposed processing block is data augmentation for enhancing COVID-19 detection by adding more data. Modern deep learning method, including CNN14, relies on large data since it tends to be more effective than small data \cite{Halevy2009,IanGoodfellowYoshuaBengio2015}. We evaluated two augmentation methods, namely SpecAugment \cite{Park2019} and noise addition using MUSAN noise corpus \cite{Snyder2015a}, and a combination of both. The result is shown in Table \ref{tab:aug_var}. Note that all methods by default utilize mixup augmentation \cite{Zhang2018} in addition to these evaluated augmentation methods.

The results show that SpecAugment is the most effective method for this cough-based COVID-19 detection task. The UA of classification with SpecAugment is 86.39\%; a large gap with the noise addition (UA = 81.4\%) and a combination of both (UA = 83.19). It has also been found that adding more data does not always improve the performance of the model, as shown in other research \cite{Atmaja2022d}. With a double number of data, a combination of SpecAugment and noise addition only improves the UA by 1.5\%, while a single SpecAugment method improves the UA by 4.7\%. The previous best result (in Table \ref{tab:seg_var}) was obtained using a combination of SpecAugment and noise addition. The next experiments are conducted to tune the hyperparameter based on this SpecAugment data augmentation method. Note that without augmentation result (along with other augmentation methods) still has a mixup augmentation method as a default configuration in the baseline method.

\begin{table}[htbp]
    \centering
    \caption{UA results on different data augmentation methods}
    \begin{tabular}{c c c}
        \hline Segmentation method & Augmentation method & UA \\
        \hline 
        \multirow{4}{*}{ Hysteresis comparator } & Without augmentation & $81.68 \%$ \\
        & SpecAugment & \textbf{86.36\%} \\
        & Noise addition & $81.40 \%$ \\
        & SpecAugment + Noise addition & $83.19\%$ \\
        \hline
        \end{tabular}
    \label{tab:aug_var}
\end{table}

\subsection{Ablation study}
In the last part of this study, we tuned three hyperparameters by using fixed values in two parameters to find an optimal value for a specific parameter. These hyperparameters optimized linearly in this independent evaluation are alpha mixup, learning rate, and weight decay. Mixup is one of the data augmentation methods that is used to generate new data from existing data. The alpha mixup is a hyperparameter that controls the amount of mixup. The learning rate is a hyperparameter that controls the step size of the gradient descent. The weight decay is a hyperparameter that controls the regularization of the model. In this study, alpha mixup is varied in a range $[0.1, 1.0]$, and learning rate and weight decay are varied in a range $[10^{-1}, 10^{-5}]$. Alpha mixup is varied with addition of $10^{-1}$. Both learning rate and weight decay are varied in a step of multiplication of $10^{-1}$. The results are shown in Table \ref{tab:ablation}.

We improve the performance of cough-based COVID-19 detection by changing the default alpha mixup \cite{Zhang2018} from 0.9 to 0.5. The default hyperparameters on previously reported UAs are 0.9, 0.001, and 0.0001 for the alpha mixup, learning rate, and weight decay, respectively. The other hyperparameters (learning rate and weight decay) are at their best. Choosing the proper parameter for mixup exhibits a potential solution to alleviate the issue of memorization and sensitivity during the training stage. The UA of this model, with 88.19\%, is the highest obtained UA in this study. 

\begin{table}[htbp]
    \centering
    \caption{UA results (\%) on ablation study with a variation of alpha mixup, learning rate, and weight decay (default value is 0.9, 0.001, and 0.01)}
        \begin{tabular}{c c | l c | l c}
            \hline 
            Alpha mixup & UA & Learning rate & UA &  Weight decay & UA  \\
            \hline 
            $0.1$ & $86.5$ & $0.1$ & $53.53$ & $0.1$ & $87.87$ \\
            $0.2$ & $86.91$ & $0.01$ & $49.17$ & \textbf{0.01} & \textbf{88.19} \\
            $0.3$ & $87.74$ & \textbf{0.001} & \textbf{88.19} & $0.001$ & $84.43$ \\
            $0.4$ & $85.12$ & $0.0001$ & $85.81$ & $0.0001$ & $86.36$ \\
            \textbf{0.5} & \textbf{88.19} & $0.00001$ & $49.41$ & $0.00001$ & $87.46$ \\
            $0.6$ & $84.84$ & & & & \\
            $0.7$ & $84.98$ & & & & \\
            $0.8$ & $86.36$ & & & & \\
            $0.9$ & $86.36$ & & & & \\
            $1.0$ & $85.81$ & & & & \\
            \hline
            \end{tabular}
    \label{tab:ablation}
\end{table}

\subsection{Summary with previous studies}
Although it is not possible to compare our results directly with previous studies due to data differences (after cough detection and segmentation), we provide a summary study shown in Table \ref{tab:summary}. The results show that our proposed method outperforms the previous studies with a large margin, although we did not utilize the ensemble method. The most similar method to ours is the method by Cassanova et al. \cite{Casanova2021}, from which our method is derived. Three processing blocks empirically improve that baseline method largely, including improvements from the original baseline methods \cite{Schuller2021}. A thorough summary without our result is also can be found in \cite{Coppock2022}, in which the authors also discuss the result of speech-based COVID-19 detection in addition to the cough-based method. 

\begin{table}[htbp]
    \caption{Summary of UA scores (\%) from different studies on the same ComParE-CCS 2021 test set; note that our method uses the same test set as others but with a different number of samples due to cough detection and segmentation methods}
    \centering\begin{tabular}{l c p{3cm} p{2.3cm} c c}
    \hline
    Reference & Data Aug. & Feature & Classifier & Ensemble & UA \\
    \hline 
    Baseline \cite{Schuller2021} & $\times$ & openSMILE, openXBOW, DeepSpctrum, AuDeep & SVM, End2You & $\surd$ & 73.9 \\
    Casanova et al. \cite{Casanova2021} & $\surd$ & log mel spectrogram & CNN14 & $\surd$ & 75.9 \\
    Illium et al. \cite{Illium2021} & $\surd$ & log mel spectrogram & Vision Transformer & $\times$ & 76.9 \\
    Solera-Urena et al. \cite{Solera-Urena2021} & $\times$ & TDNN-F, VGGish, PASE+ & SVM & $\surd$ & 69.3 \\
    Ours & $\surd$ & log mel spectrogram & CNN14 & $\times$ & \textbf{88.19} \\
    \hline
    \end{tabular}
    \label{tab:summary}
\end{table}









\section{Conclusions}
In this paper, we evaluated three important processing blocks for training cough sounds for COVID-19 detection. These blocks are cough detection, cough segmentation, and data augmentation. Although the last block is common in machine learning and deep learning techniques, the first two blocks are distinctive to cough-based COVID-19 detection. The gains using these blocks are 2.86\%, 7.65\%, and 3.17\%, respectively, relative to the addition of one block after others. These gains were obtained using cough detection rate at 90\%, cough segmentation using hysteresis comparator, and data augmentation using SpecAugment. Furthermore, we optimize the hyperparameters through an ablation study by tuning alpha mixup, learning rate, and weight decay. A proper value of alpha mixup (alpha=0.5) improves the UA 88.19\%, which is the highest UA in this study. The learning rate and weight decay are at their best since the baseline method (learning rate=0.001 and weight decay = 0.01). The results show that the proposed method outperforms the previous studies with a large margin, although we did not utilize the ensemble method.

Future studies should be directed to test this research-stage method for preliminary COVID-19 detection in real world. The results should be justified by accurate COVID-19 labeling, e.g., by PCR test. Explainable AI techniques can be used to explain and understand the model's decision, which is important for the clinical use of AI-based COVID-19 detection.

\section{Acknowledgements}
This paper is partly based on results obtained from a project, number 1014/PKS/ITS/2022, funded by Directorate of Research and Community Service, Sepuluh Nopember Institute of Technology (ITS), Indonesia. The authors would like to thank to Dr. Dhany Arifianto of VibrasticLab ITS for allowing us to use his computational resources for this study.

\bibliographystyle{elsarticle-num-names} 
\bibliography{covid-19}
\end{document}